\documentclass[conference,a4paper]{IEEEtran}
\usepackage{xcolor}
\ifCLASSINFOpdf
 \usepackage[pdftex]{graphicx}
\else
\usepackage[dvips]{graphicx}
\fi
\ifCLASSOPTIONcompsoc
 \usepackage[caption=false,font=normalsize,labelfont=sf,textfont=sf]{subfig}
\else
 \usepackage[caption=false,font=footnotesize]{subfig}
\fi
\usepackage{siunitx}
\usepackage{url}
\usepackage{amsthm}
\usepackage{mathtools}
\usepackage{array}
\usepackage{cite}
\usepackage{tabularx}
\usepackage{amsmath,amssymb,amsfonts}
\usepackage{algorithmic}
\usepackage{graphicx}
\usepackage{textcomp}
\usepackage{bm}
\usepackage{siunitx}
\usepackage{dirtytalk}
 \usepackage{multicol}
\usepackage{breqn}
\begin{document}

\title{A Near-Field Focused Phased-Array Antenna Design Using the Time-Reversal Concept for Weed Control Purpose
}
\author{\IEEEauthorblockN{
Adel Omrani\IEEEauthorrefmark{1}  
Guido Link\IEEEauthorrefmark{1},      
and John Jelonnek\IEEEauthorrefmark{1}\IEEEauthorrefmark{2} 
}                                 
\IEEEauthorblockA{\IEEEauthorrefmark{1}IHM, Karlsruhe Institute of Technology (KIT), Karlsruhe, Germany} \IEEEauthorblockA{\IEEEauthorrefmark{2}IHE, Karlsruhe Institute of Technology (KIT), Karlsruhe, Germany}
Email:  adel.hamzekalaei@kit.edu}
\maketitle
\begin{abstract}
Near-field focus (NFF) antennas have been recently used in several applications for different purposes. In this work, the time-reversal (TR) concept is used to shape the phase distribution of the phased array elements for the NFF of the electromagnetic field strength. It is shown that the TR concept is equivalent to the known ray optic method for the NFF.  A slotted waveguide phased array antenna operating at \textbf{5.8\ GHz} is designed to provide the maximum electric field strength at the near-field region of the phased array. It is shown that the application of the full-wave simulation allows for an antenna design that provide high strength of the electromagnetic field and sufficient steerability even at near-field conditions close to the phased array antenna.
 \end{abstract}

\vskip0.5\baselineskip

\section{Introduction}
In farming production, weeds compete with crops for sunlight, space, nutrients, water, and CO$_2$ and can significantly impact crop products worldwide. Dense weed growth can make harvesting very difficult and reduce it significantly. Even though an estimated 3 billion kg of pesticides is currently applied worldwide, it is estimated that $37\ \%$ of global crop production is still lost. $13\ \%$ of this is due to insects, $12\%$ to plant pathogens, and $12\ \%$ to weeds \cite{Chem}.\newline
\indent Controlling and demolition of the distribution of weeds in a crop field is vital to increase the production rate \cite{Graham1, Graham2}.  Weed control by microwaves is an environmentally friendly method for replacing chemical or pesticide methods, which are no longer acceptable from an ecological point-of-view. One novel idea is to utilize a phased array antenna to concentrate the electromagnetic (EM) power at the area of the weed. In this regard, it's desired i) to achieve a high power density at the desired area and ii) to allow the focus of the microwave beam to be quickly and specifically targeted on the previously identified weed's location.\newline
Delivering the maximum possible EM power at the expected area of the weed demands a near-field focused (NFF) phased array antenna. NFF antennas have been employed in various areas like industrial microwave applications, wireless power transfer, RFID applications, etc. The idea of the NFF is to contribute the phases of the array's elements in such a way as to have a focal point in the phased array's near-field region.\newline
To determine the phase distribution of the phased array antennas' elements for the NFF, a ray optic concept is proposed that correctly considers the phase difference between the far-field (FF) region and the near-field (NF) region. The idea comes from the point that in the NF region, the distance between two adjacent elements significantly affects the phase value and should be correctly considered. The details of the method are described in \cite{NFF}.\newline
In this work, we use a time reversal (TR) concept to obtain the correct phase distribution of each element of the designed phased array antenna while performing in free space. The employed concept can be extended when an air-soil interface is located near a phased array antenna while the ray optic concept cannot. in Section \ref{sec:Theory}, the concept of the TR is addressed. Section \ref{sec:Design} discusses the antenna design and configuration. Finally, the simulation results and discussion is provided in Section \ref{Sec:Simulation}, and Section \ref{Conclusion}, respectively. 
\begin{figure}[!b]
\centering
 \includegraphics[width=0.22\textwidth]{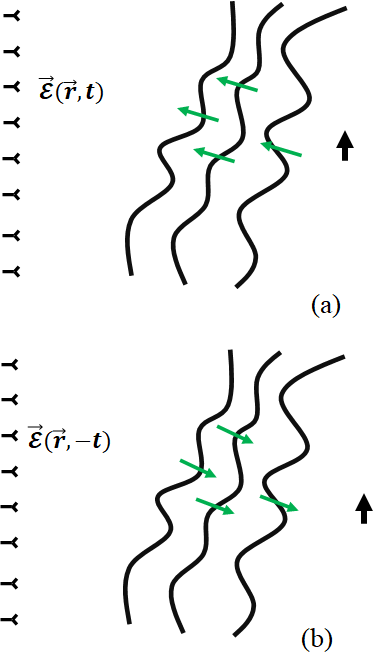}
\caption{(a) Radiation of EM fields using an infinitesimal dipole antenna, (b) Radiation of the time-reversed EM fields.}
   \label{F1}
 \end{figure}
\section{Near-Field Focused Theory}\label{sec:Theory}
As mentioned earlier, despite the far filed region of an antenna, the phase difference in the near-field antenna cannot be suppressed and may have a significant effect on the distance of the focusing point. In order to compensate for the phase difference at the NFF region, one approach employs the ray optics approximation. Another way to obtain the exact phase of the radiating elements is by using Green's function and the TR concept.\newline
The TR concept was first introduced in ultrasonic to locate the position of inhomogeneity in an interested media\cite{Fink1, Fink3}. In a non-dispersive media, the wave equation is time-symmetric. This implies both $\vec{\mathcal{E}}(\vec{r},t)$ and $\vec{\mathcal{E}}(\vec{r},-t)$ are the solution of the following wave equation
\begin{equation}
    \nabla^{2}\Vec{\mathcal{E}}(\Vec{r},t)-\mu(\Vec{r})\varepsilon(\Vec{r})\frac{\partial^2}{\partial t^2} \Vec{\mathcal{E}}(\Vec{r},t)=0
\end{equation}
where $\mu$  and $\varepsilon$ denote the permeability and permittivity of the medium, respectively. $\Vec{r}$ denotes any point in the media, and $t$ represents the time. So, this guarantees for every wave diverging away from the source, there exists a reversed wave that would precisely retrace the path of the original wave back to the source.\newline
\indent To experimentally examine the TR method, an excitation signal was sent to the region of interest, and the reflected signal was recorded using the transducers surrounding the media. The received signal was time reversed and used as a new excitation for the transducers. Later, this new excitation (time-reversed signal) was retransmitted to the region of interest, and it was observed that the newly sent signals focused on the target (inhomogeneity). It indicates that the converged time-reversed signals travel the same path as the diverged signal travels from the inhomogeneity\cite{Fink2}.\newline
\indent Here, we use an infinitesimal dipole antenna (or point source) positioned at the desired focal $\vec{r}_{s}=(x_s,y_s.z_s)$ to excite the media, as shown in Fig. \ref{F1} (a). This can be interpreted as a secondary source that generates the diverged scattered fields represented by $\Vec{\mathcal{E}}(\Vec{r},t)\leftrightarrow E(\Vec{r},f)$, where $f$ is the frequency. Now, the TR concept implies there exist converged waves (represented by $\Vec{\mathcal{E}}(\Vec{r},-t) \leftrightarrow E^{*}(\Vec{r},f)$) ($*$ is the complex conjugate operation) that retrace the same path and focus on the location of the original source. Hence, the complex electric field at the location of the radiating elements will be obtained. Instead of time-reversing the electric field, it's equivalent in the frequency domain, i.e., a conjugate operation is applied. This new set of obtained phases will be employed as an excitation for the radiating elements to provide a focus at the desired point.\newline
Assume that the antenna array is located in the free space. To obtain the proper phase of the antenna, we placed an infinitesimal dipole antenna at the location of the desired focal point. The electric field at the location of the antenna array will be obtained as follows 
\begin{equation}
\begin{split}
    E(\Vec{r}_{a_{n}},f)=j\omega \mu_{0}G_{e}(k_{0},\Vec{r}_{a_{n}},\vec{r}_{s})&\\=j\omega \mu_{0}\frac{e^{-jk_{0}|\Vec{r}_{a_{n}}-\vec{r}_{s}|}}{4\pi |\Vec{r}_{a_{n}}-\vec{r}_{s}|}
    \end{split}
\end{equation}
for $n=1,2,...,N$, where $k_0$ is the wavenumber of free space, $\omega$ is the angular frequency, and $G_{e}$ denotes free space electric Green's function.  Now, by conjugating the received electric field, the proper phase of the $n^{th}$ element of the phased array antenna will obtain as follow
\begin{equation}
    \varphi_{n}=\angle (E^{*}_{n}(\Vec{r}_{a_{n}},f)).
\end{equation}
This new set of phases will be used to provide an NFF at the desired area.
\section{Phased Array Antenna Configuration and Design}\label{sec:Design}
A slotted waveguide (SWA) phased array antenna is proposed and designed to demonstrate the feasibility of weed control using the phased array antenna in a radiative near-field domain. The SWA (i) is appropriate for high-power applications, (ii) enables the fabrication of the phased array with a low number of elements, and (iii) provides the required focused area in both longitudinal and lateral directions. Due to the distribution of the slots in the longitudinal direction, the radiation pattern in H-plane is already focused and in the E-plane it is not. Also, by controlling the number of slots, the size of the focal point in this direction can be adjusted. Needless to say, no more antennas are required in this direction, which can reduce the cost and complexity of the phased array antenna and overall system. By increasing the number of antennas, the focus E-plane can be achieved.\newline
\begin{figure}[!t]
\centering
\includegraphics[width=0.40\textwidth]{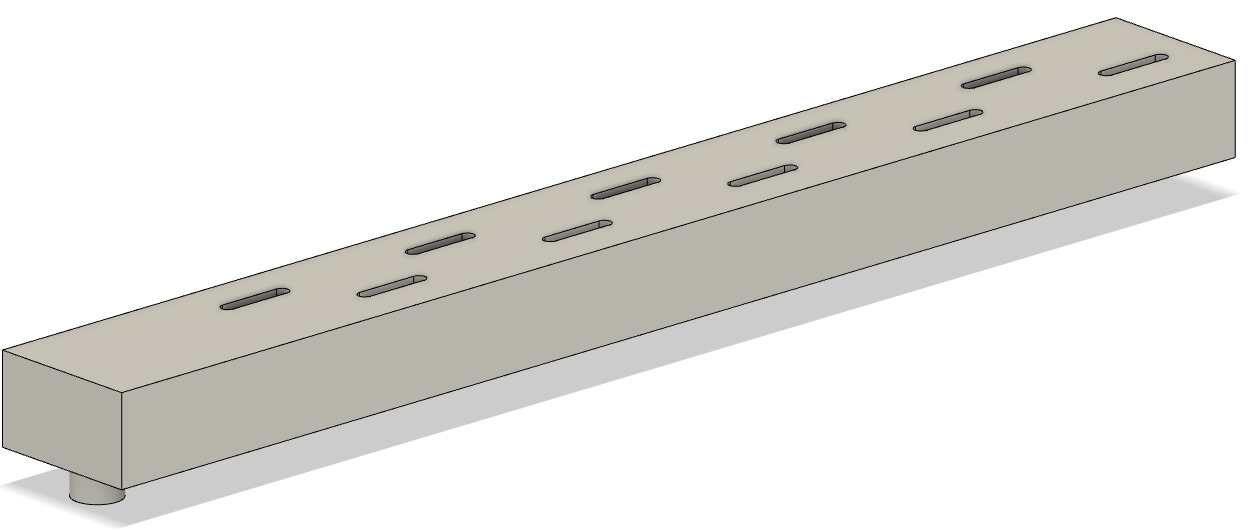}
\includegraphics[width=0.5\textwidth]{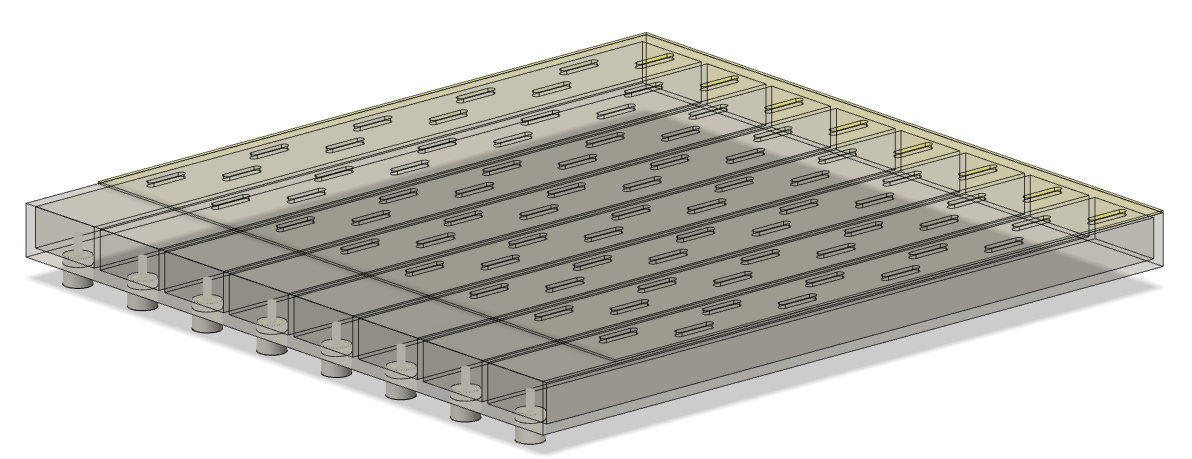}  
\caption{Schematic of the SWA antenna (top) top view, (bottom) phased array.}
   \label{STr}
 \end{figure}

\subsection{SWA Design parameters}
\begin{figure}[!t]
\centering
 \includegraphics[width=0.5\textwidth]{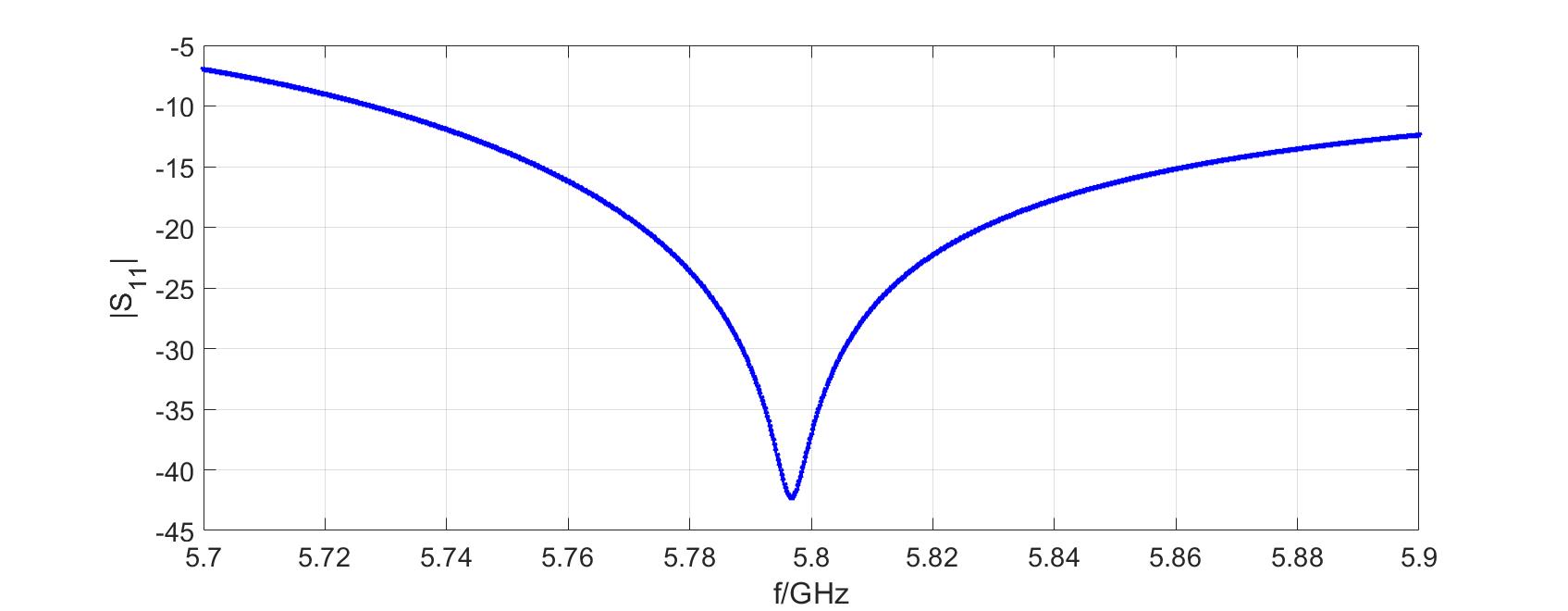}
\caption{Return loss in dB of the single SWA.}
   \label{S11}
 \end{figure}

Fig. \ref{STr} (top) and Fig. \ref{STr} (bottom) give the schematic of the single SWA and proposed SWA phased array antenna, which operates at \SI{5.8}{GHz}. The dimension of the cross-section of the waveguide is \SI{40.4}{mm}$\times$\SI{19.8}{mm}. Each SWA contains $10$ slots with a length of \SI{22.4}{mm} and a width of \SI{4}{mm}. For ease of fabrication, the slots are slightly curved. The center distance of the slots to the side wall is \SI{11}{mm}. The center-to-center distance of two slots in the longitudinal direction is \SI{32}{mm}. The distance between the last slots to the short wall is \SI{10}{mm}. Moreover, a Radom is considered in the design process to preserve the array against dust and drops. To do this, a PTFE layer with a thickness of \SI{1}{mm} was chosen. A coaxial to waveguide transition is separately designed and unified with the SWA antenna. In this regard, a standard N-type is designed and integrated into the coaxial to waveguide transition for the high-power application purpose of the antenna. The simulated return loss of the designed SWA is depicted in Fig. \ref{S11} in the ISM frequency range.
\section{Simulation Results}\label{Sec:Simulation}
\subsection{Time-Reversal Concept and Ray Optic Comparison}
\begin{figure}[!b]
\centering
 \includegraphics[width=0.5\textwidth]{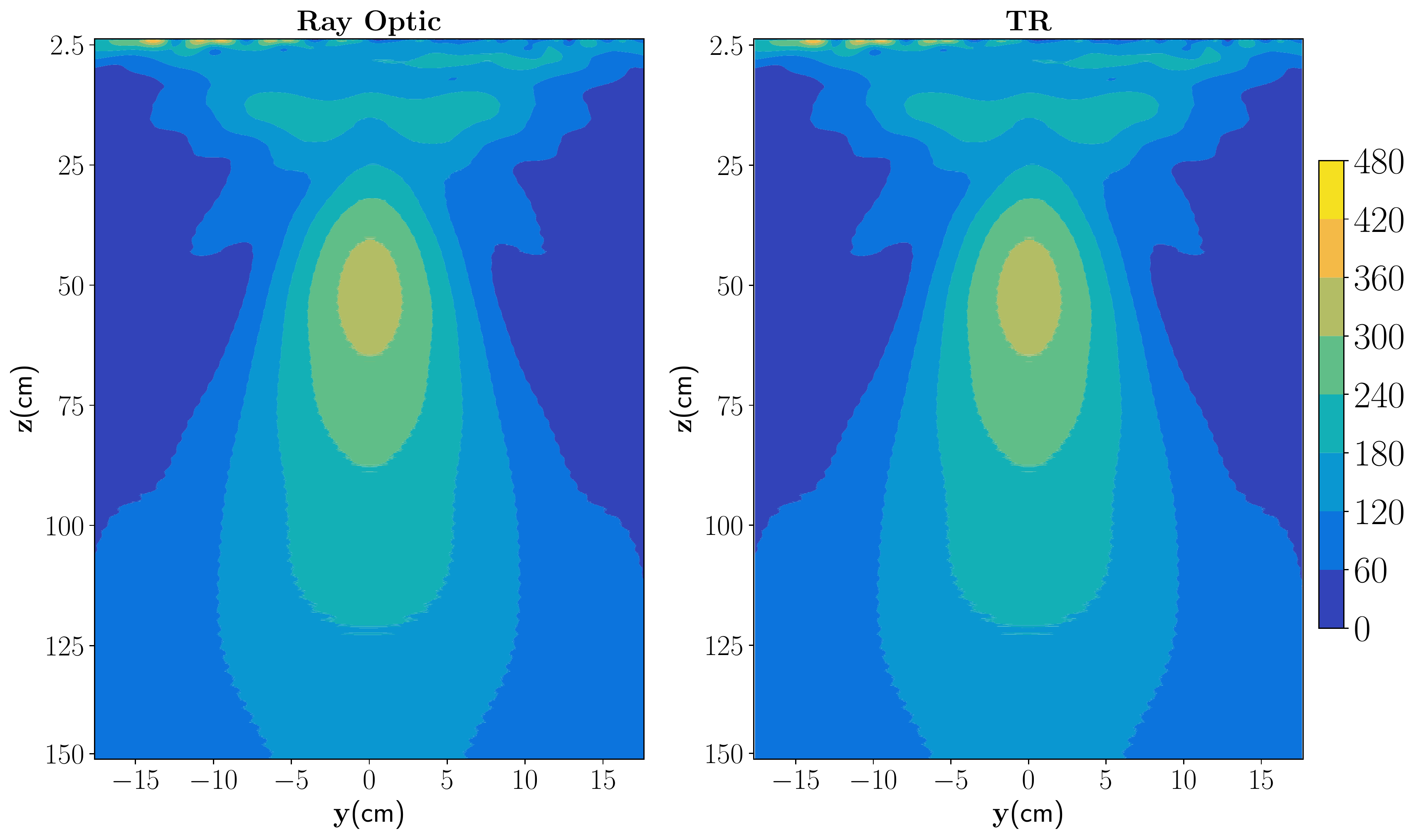}
\caption{Comparison between the ray optic and TR method.}
   \label{raytr}
 \end{figure}
A full wave simulation using commercial software CST Studio Suite is fulfilled to obtain the electromagnetic field distribution in the interested domain. Fig. \ref{raytr} represents the comparison between the simulated electric field using the ray optic and the proposed TR method. As can be seen from this figure, the TR approach provides the same result as the ray optic technique. However, it's worth mentioning that these two methods' excitation phases of the phased array elements are different. 
\begin{figure}[!t]
\centering
 \includegraphics[width=0.5\textwidth]{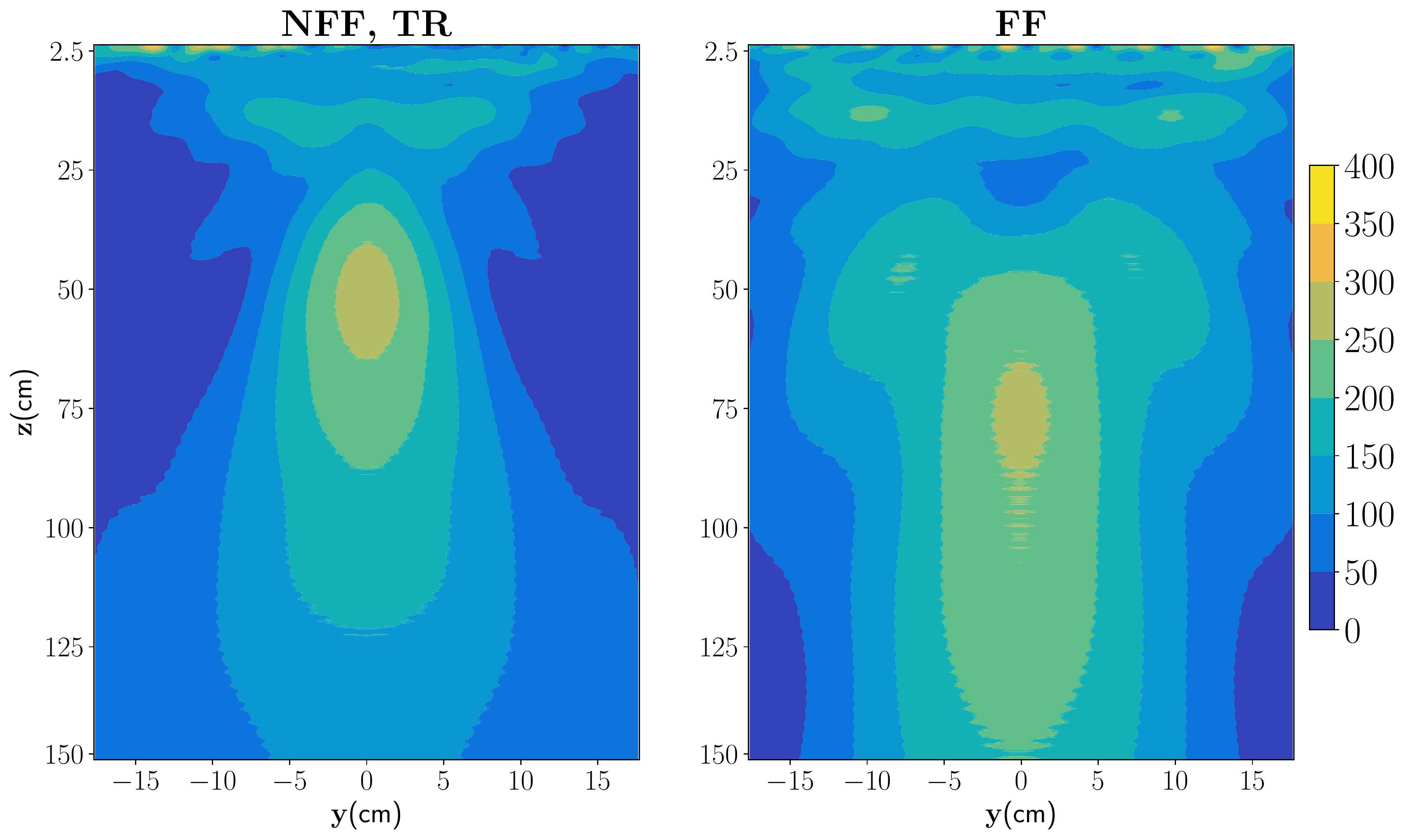}
\caption{Comparison between the TR method and far-field phase method.}
   \label{NFFFF}
 \end{figure}
\subsection{Comparison Between Far Field Phase Method and TR Concept}
\begin{figure}[!t]
\centering
\includegraphics[width=0.5\textwidth]{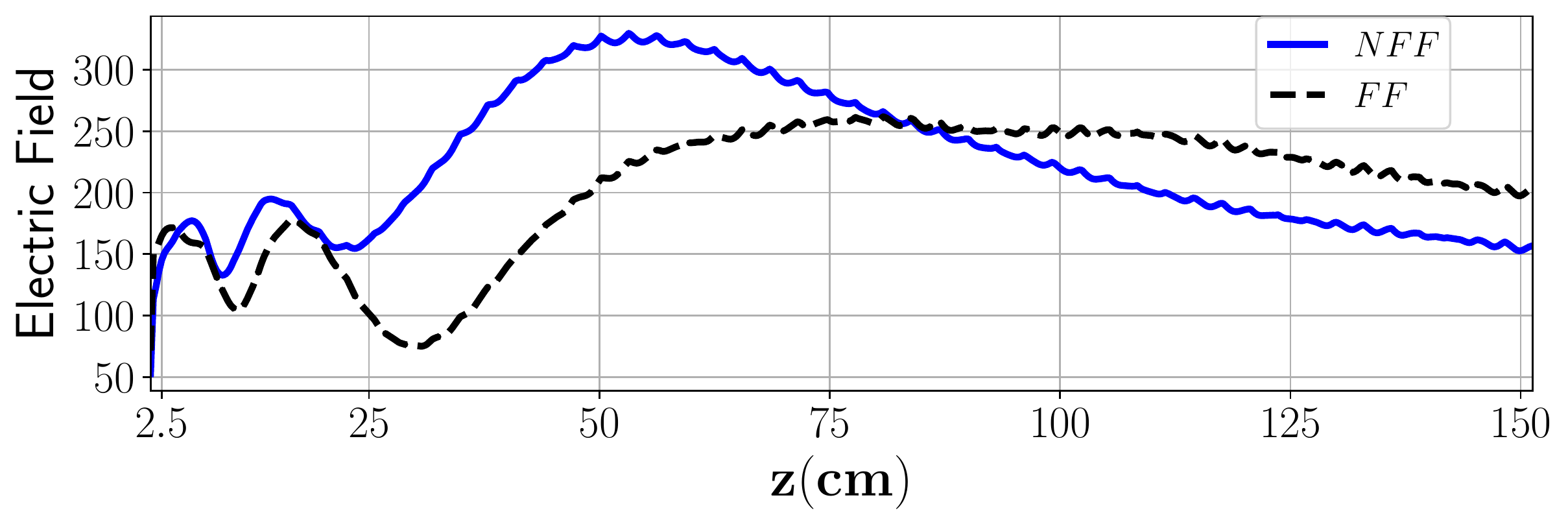}
\caption{Comparison of E-field profile in z-direction in the E-plane}
\label{Efield}
\end{figure}

Figure \ref{NFFFF} represents the electric field distribution when the far field assumption is used to shape the phase of the radiating elements. As can be understood from this figure, (i) the focused point is occurring at a further distance from the phased array, and (ii) the maximum electric field of the focused point is less than the TR method. For a better comparison, the electric field distribution versus the axial distance of the antenna is plotted for two methods and compared in Fig. \ref{Efield}. As can be understood by this figure, compared to the FF assumption, employing the NFF phase provides a higher electromagnetic field strength at a shorter distance.
\subsection{Steerability of the Phased Array}
Different associated phase distributions are obtained using the TR concept to evaluate the steerability of the phased array. A dipole is inserted at the desired focused point, and the electric field is calculated at the position of the phased array elements. Then by conjugating the phase of the electric field, a new set of excitation is obtained for the phased array antenna.
 \begin{figure}[!hbt]
\centering
 \includegraphics[width=0.5\textwidth]{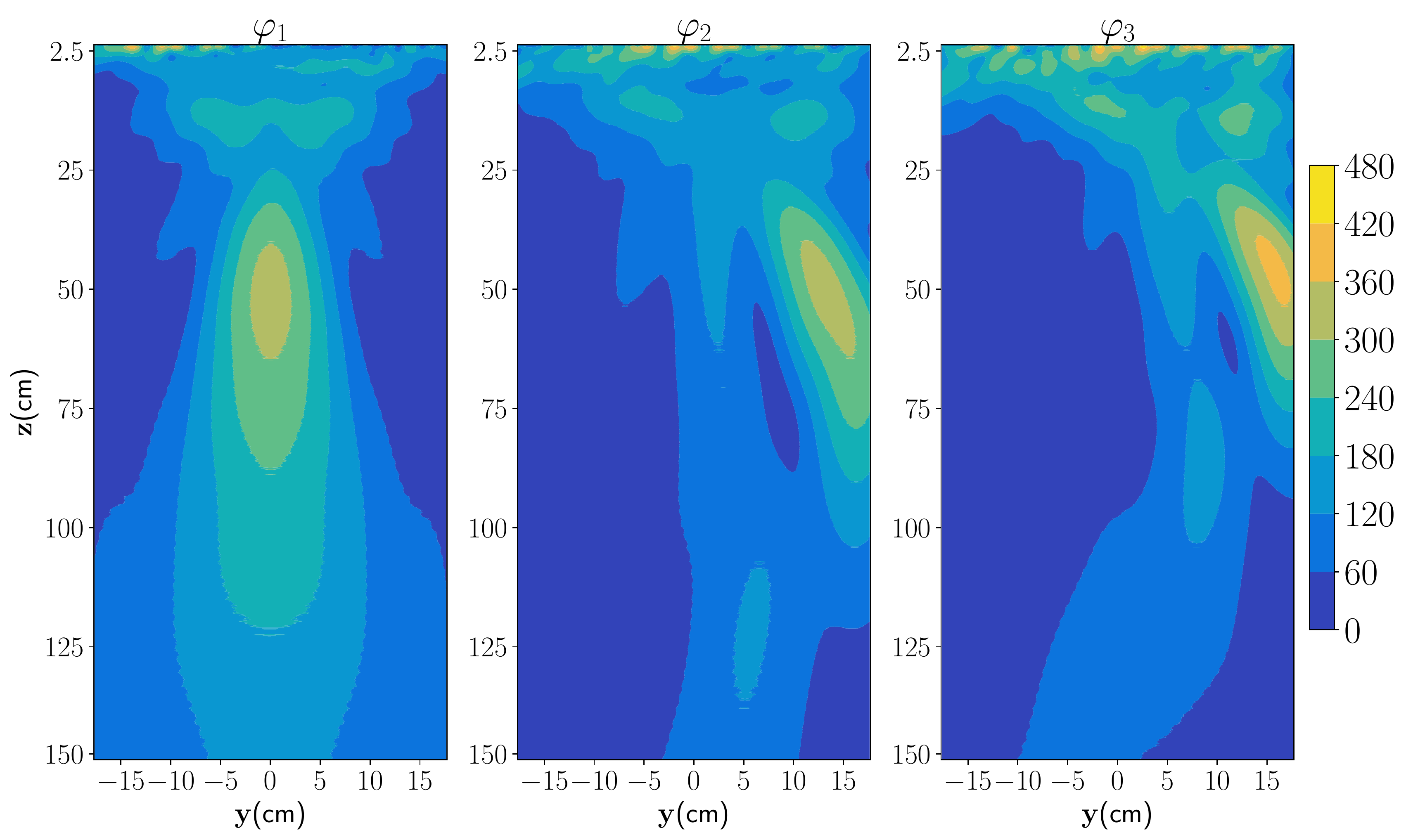}
\caption{Comparison between the ray optic and TR method.}
   \label{steer}
 \end{figure}
 \section{Conclusion and Future Works}\label{Conclusion}
 In this work, a \SI{5.8}{GHz} SWA phased array antenna was designed for weed control purposes. A TR concept was used and investigated for NFF that can increase the concentrated power in the desired area. It was shown that the TR concept can provide the same results as the ray optic method in free space. The desired steerability for the NFF phased array can be achieved using the TR concept. However, in future work, by considering the air-soil interface near a phased array antenna, it will be shown that the performance of the ray optic method will be rendered while TR provides a proper phase distribution.

 \section{Acknowledgment}
 This Project is supported by the Federal Ministry for Economic Affairs and Climate Action (BMWK) on the basis of a decision by the German Bundestag. (project number KK5431301DF1)
\IEEEtriggercmd{}
\IEEEtriggeratref{4}

\bibliography{references}

\end{document}